\documentstyle[sprocl]{article}
\begin{document}
 \title{ SUPERCURRENT DECAY BY VORTEX NUCLEATION IN THE PRESENCE OF 
AN ARRAY OF PINNING CENTRES: A RENORMALIZATION GROUP APPROACH 
}
 \author{ ROBERTO IENGO  }
 \address{ International School for Advanced Studies, Via Beirut 4, 
34014 Trieste (Italy) 
   }
 \author{ GIANCARLO JUG  }
 \address{ Dipartimento di Fisica, Universit\`a di Milano, Via 
Lucini 3, 22100 Como (Italy) \\
and Max-Planck-Institut f\"ur Physik Komplexer Systeme, Au\ss enstelle
Stuttgart, Postfach 800665, D-70506 Stuttgart (Germany) 
   }
 \maketitle\abstracts{ We study the phenomenon of decay of a 
supercurrent in a superconducting thin film due to the spontaneous 
homogeneous nucleation of quantized vortex-antivortex pairs in the 
absence of an external magnetic field and in the presence of a
periodic pinning potential. We describe the vortex nucleation by means 
of a Schwinger-type path-integral calculation including the effects of 
quantum dissipation. The overdamped case is treated exactly, whilst 
the case in which a periodic array of pinning centres is present is      
dealt with by means of a renormalization group (RG) approach in frequency 
space. This yields an approximate but very appealing analytic result for
the dependence of the vortex nucleation rate on the externally-driven
supercurrent. In this formula the rate dependence displays oscillations 
which are connected to the pinning periodicity and correspond to the
vortex nucleation length probing over the pinning sites as the current 
density is varied. Our RG treatment describes the vortex nucleation 
process in both the confined and the mobile phases of the 
dissipation-driven localization transition characterising the motion of
dissipating quantum particles in a periodic potential. }

 \section{ Introduction }

There has been a strong revival of interest, recently, in the physics 
of magnetic vortices in type II and high-temperature superconductors 
\cite{reviews}. Most research efforts have been devoted to phenomena 
relating to the nature of the mixed phase of a superconductor in some 
externally applied magnetic field and supercurrent. Issues connected
with the pinning of the flux lines by defects have been widly studied. 

We \cite{ieju}, as well as Ao and Thouless \cite{aoth} and Stephen
\cite{stephen}, have addressed the problem of the quantum dynamics of
vortices in the absence of an external field but in the presence of
an externally driven supercurrent, quantum dissipation and pinning.
This leads to the decay of a supercurrent, or a residual zero-temperature
resistance in the superconductor. Whilst most of the dissipation seems
to be ascribed to vortices tunneling in the sample from the edge, an
interesting novel possibility also explored by us in depth is that of
a residual resistance arising from spontaneous vortex-antivortex pair
creation in the bulk of a thin film. This is the mesoscopic counterpart 
of electron-positron pair production of two-dimensional (2D) quantum 
electrodynamics (QED) in the presence of static e.m. fields, which in a 
superconductor arise from the static and velocity-dependent components of 
the Magnus force acting on the vortices. Exploiting this analogy with QED, 
a powerful ``relativistic'' quantum field theory approach has been 
set up to study vortex nucleation in the 2D geometry in the presence of 
quantum dissipation and of pinning potentials. The central result is that 
the nucleation rate $\Gamma$ has a strong exponential dependence on the 
number current density $J$, given by

\begin{equation} 
\Gamma{\propto}\eta^{1/2}\eta_{eff}J^{-1}
\exp\{-\eta_{eff}{\cal E}_{0R}^2/4\pi J^2\}
\label{rate0}
\end{equation}

\noindent
Here $\eta_{eff}$ is an effective viscosity coefficient as renormalised by 
the magnetic-like part of the Magnus force, and ${\cal E}_{0R}$ is the rest-
or nucleation-energy of a single vortex as renormalized by screened
Coulomb interactions and (fake) Landau-level corrections. This 
exponential dependence would make the vortex nucleation (folded, e.g.,
into the sample's resistance) observable in a rather narrow range of 
$J$-values. Thus, normally the superconductor is essentially resistance-free.
However, the high values of $J$ that can be reached in the high-$T_c$
materials make the possibility of observing pair creation in static fields 
within reach for thin films. One particular feature that would uniquely 
relate the residual resistance to the phenomenon of spontaneous vortex-pair 
creation is the presence of {\em oscillations} in the $J$-dependence of 
$\Gamma(J)$ in case a {\em periodic} pinning potential is artificially 
created in the film. These oscillations are in fact strictly connected to 
the pinning-lattice spacing $d=2\pi/k$ of the periodic potential (we assume 
a square lattice), e.g.

\begin{equation}
U({\bf q}(t))=U_0 \sum_{a=1}^2 \left [ 1 - \cos \left ( kq_a(t) 
\right ) \right ]
\label{potent}
\end{equation}
   
\noindent
acting on the nucleating vortex-pairs described by a coordinate ${\bf q}$. 

The problem of quantum dissipation for a particle moving in a periodic
potential has some interesting features in its own right 
\cite{schmid,ghm,fizw}. It is characterised by a localization phase
transition driven by dissipation; accordingly, two phases can occur
depending on whether the dissipation coefficient \cite{cale} $\eta$ is
greater (confined phase) or smaller (mobile phase) than a critical
value $\eta_c=k^2/2\pi=2\pi/d^2$. This localization transition is described
by a Kosterlitz-type renormalization group (RG) approach, yet with some
important differences that will be recalled below. We have implemented 
the RG approach for the evaluation of the dependence of the spontaneous
nucleation rate of vortex-antivortex pairs on the external parameters for 
our own quantum dynamical system. A remnant of the dissipation-driven
phase transition is observed and the pair production rate $\Gamma$ can
be derived in both phases by means of a frequency-space RG procedure  
leading to observable current-oscillations if $\eta > \eta_c$.

 \section{ RG approach to dissipative localization transition } 

First, we briefly recall the RG description of the localization
transition driven by quantum dissipation \cite{fizw}. The effective 
action for a particle diffusing in a periodic potential and subject to 
quantum dissipation of the Caldeira-Leggett type \cite{cale} is, in 
Fourier frequency space:

\begin{equation}
{\cal S}=\int_0^{\tau}{\cal L}({\bf q})=\tau
\sum_n \{ \frac{1}{2}m\omega_n^2+\frac{1}{2}\eta |\omega_n| \}
\bar{q}_a(\omega_n)\bar{q}_a(-\omega_n)+\int_0^{\tau} dt U({\bf q})
\label{action0}
\end{equation}
 
\noindent
where $m$ is the mass of the quantum particle and $\eta$ the 
phenomenological friction coefficient. In the low-frequency limit the 
effects of inertia can be neglected and the problem would acquire the same 
phenomenology as for the sine-Gordon model (in (0+1)-dimensions), 
except for the peculiar $\omega_n$-dependence of the propagator reflecting 
the broken time-reversal symmetry of quantum dissipation. When the RG 
procedure is applied to Eq. (\ref{action0}) a renormalization of the 
potential amplitude $U_0$ occurs, but not of the friction coefficient 
$\eta$ since only local operators in the time variable can be generated 
within a RG transformation. In terms of the dimensionless parameters 
($\Omega$ is a large frequency-cutoff) ${\cal U}=U_0/\Omega$ and 
$\alpha=2\pi\eta/k^2$, the RG recursion relations read

\begin{equation}
\frac{d{\cal U}}{d\ell}=\left ( 1-\frac{1}{\alpha} \right ) {\cal U} 
+ \cdots, \qquad
\frac{d\alpha}{d\ell}=0
\label{recrel}
\end{equation}

\noindent
with $e^{-\ell}$ the frequency-scale renormalization parameter. These have 
the simple solution

\begin{equation}
{\cal U}(\ell)={\cal U}(0)e^{(1-\eta_c/\eta)\ell}, \qquad  
\alpha(\ell)=\alpha(0)
\label{rgflow}
\end{equation}

\noindent
displaying the localization transition for $\eta=\eta_c=k^2/2\pi=2\pi/d^2$.
The potential's amplitude vanishes
under a RG change of time scale for $\eta < \eta_c$, but for  
$\eta > \eta_c$ it tends to diverge and the RG procedure must be 
interrupted. Unlike in the Kosterlitz RG scheme, this cannot be done 
unequivocally in the present situation, for there is no true characteristic 
correlation time or frequency owing to the fact that one never moves away 
from the neighbourhood of the critical point $\eta_c$. An alternative 
strategy for the confined phase is to resort to a variational treatment
\cite{fizw}, which dynamically generates a correlation time.
In this procedure the action of Eq. (\ref{action0}) is replaced by a
trial Gaussian form (neglecting inertia)

\begin{equation}    
{\cal S}_{tr}=\frac{\eta}{4\pi} \int_0^{\tau} dt \int_{-\infty}^{+\infty} dt'
\left ( \frac{{\bf q}(t)-{\bf q}(t')}{t-t'} \right )^2 + \frac{1}{2} M^2 
\int_0^{\tau} dt {\bf q}(t)^2
\label{actiontr}
\end{equation}

\noindent
where $M^2$ is determined by minimising self-consistently the free energy
$F_{tr}+\langle S-S_{tr} \rangle_{tr}$. This leads to the equation

\begin{equation}
M^2=U_0k^2 \exp \left \{ -\frac{k^2}{2\tau} \sum_n \frac{1}{\eta|\omega_n|
+M^2} \right  \}
\end{equation}

\noindent
having a solution $M^2{\neq}0$ only in the confined phase ($\eta > \eta_c$),
since introducing the cutoff $\Omega$ in the (continuous) sum over frequency 
modes $\omega_n=2{\pi}n/\tau$, the equation for $M^2$ leads to (for
$M^2\rightarrow 0$)

\begin{equation}
M^2=\eta\Omega \left ( \frac{2\pi U_0}{\Omega} \frac{\eta_c}{\eta} 
\right )^{\eta/(\eta-\eta_c)}{\equiv}\eta\Omega\mu
\label{mass}
\end{equation}

\noindent
This spontaneously generated ``mass'' interrupts the divergent 
renormalization of the periodic potential amplitude $U_0$, which in the 
RG limit $\ell{\rightarrow}{\infty}$ tends to 

\begin{equation}
U_0(\ell)=U_0 \left ( \frac{e^{-\ell}+\mu}{1+\mu} \right )^{\eta_c/\eta}
{\rightarrow}U_0 \left ( \frac{\mu + 1/n^{*}}
{\mu + 1} \right )^{\eta_c/\eta}
\end{equation}

\noindent
Here, we have put $\Omega=2\pi n^{*}/\tau=n^{*}\omega_1$ and 
$\mu=M^2/\Omega\eta$.

 \section{ RG treatment of vortex-antivortex pair-creation in the presence
of a periodic pinning potential }

We begin by recalling the need for a relativistic description of the
process. This leads \cite{ieju} to a Schwinger-type formula for the decay 
of the ``vacuum'', represented by a thin superconducting film in which static
e.m.-like fields arise when a supercurrent is switched on. The quantum 
fluctuations of these fields are vortex-antivortex pairs, nucleating at a
rate given by     

\begin{equation}
\frac{\Gamma}{L^2}=\frac{2}{L^2T} Im \int_{\epsilon}^{\infty}
\frac{d\tau}{\tau} e^{-{\cal E}_0^2\tau} \int
{\cal D}q(t) \exp\{ -\int_0^{\tau} dt {\cal L}_E \}
\label{rate}
\end{equation}
                  
\noindent
where $L^2T$ is the space-time volume of the sample and ${\cal E}_0$ the
vortex-nucleation energy (suitably renormalised by vortex-screening effects).
Also        

\begin{eqnarray}
{\cal L}_E&=&\frac{1}{2}m_{\mu}\dot{q}_{\mu}\dot{q}_{\mu}-\frac{1}{2}i
\dot{q}_{\mu}F_{\mu\nu}q_{\nu} + V({\bf q}) \nonumber \\
&+&\sum_k \left \{ \frac{1}{2}m_k\dot{\bf x}_k^2
+\frac{1}{2}m_k\omega_k^2 \left( {\bf x}_k+\frac{c_k}{m_k\omega_k^2}{\bf q}
\right )^2 \right \}
\label{lagran}
\end{eqnarray}
                      
\noindent
is the Euclidean single-particle relativistic Lagrangian, incorporating the
pinning potential $V({\bf q})=2{\cal E}_0U({\bf q})$ and the Caldeira-Leggett
mechanism \cite{cale}. In the absence of the pinning potential, the 
relativistic action is quadratic and the path integral in Eq. (\ref{rate}) 
can be evaluated exactly. The leading term in the expression for $\Gamma$ 
follows from the lowest pole in the $\tau$-integral and this can be obtained 
exactly in the (non-relativistic) limit in which 
$m_1=m_2=\frac{\gamma}{2}{\rightarrow}0$, with 
$\frac{1}{\gamma}={\cal E}_0/m{\rightarrow}{\infty}$ playing the role of the
square of the speed of light. The result \cite{ieju} is Eq. (\ref{rate0}).

We now come to the evaluation of $\Gamma$ in the presence of the periodic 
potential, which calls for the RG approach of Section 2. Integrating out the 
Euclidean ``time''-like component $q_3(t)$, we reach a formulation in which 
the electric-like and the magnetic-like Magnus field components are 
disentangled. In terms of Fourier components, dropping the magnetic-like part
and for $\gamma{\rightarrow}0$:

\begin{equation}
\int_0^{\tau} dt {\cal L}_E({\bf q})=\tau\sum_{n\neq 0} \{ \frac{1}{2}\eta 
|\omega_n| - E^2{\delta}_{a1} \} \bar{q}_a(\omega_n) \bar{q}_a(-\omega_n) 
+\int_0^{\tau} dt V({\bf q}) 
\label{lagranr}
\end{equation}

\noindent
with $E=2\pi J$ the electric-like field due to the supercurrent donsity $J$.
We have shown \cite{ieju} that the only role of the magnetic-like field is 
to renormalize the nucleation energy and the friction coefficient, hence our 
problem amounts to an effective one-dimensional system in the presence of 
${\bf E}$ and dissipation. The evaluation of the Feynman Path Integral (FPI)
proceeds by means of integrating out the zero-mode, $\bar{q}_0$, as well as 
the high-frequency modes $\bar{q}_n$ with $n>1$, since again the leading 
term for $\Gamma$ in Eq. (\ref{rate}) comes from the divergence of the FPI 
associated with the lowest mode coupling to ${\bf E}$. The effect of 
$\bar{q}_n$ with $n > 1$ is taken into account through the frequency-shell 
RG method of Section 2, leading to a renormalization of the amplitude 
$V_0=2{\cal E}_0U_0$ of the (relativistic) pinning potential. The 
renormalization has to be carried out from the outer shell of radius $\Omega$
to $\omega_1=2\pi/\tau$. In the mobile phase ($\eta < \eta_c$) this implies
$e^{\ell}=\Omega\tau/2\pi=n^{*}$ in Eq. (\ref{rgflow}), with (from the leading 
pole of the FPI) $\tau=\pi\eta/E^2$ (${\rightarrow}\infty$ for relatively 
weak currents). In the more interesting confined phase ($\eta > \eta_c$) we 
must integrate out the massive $n > 1$ modes with a Lagrangian
${\cal L}(\bar{q}_n)=\tau \left ( \frac{1}{2}\eta |\omega_n|+\frac{1}{2} 
M^2-E^2 \right ) \bar{q}_n\bar{q}_n^{*}$. This leads to an additional, 
entropy-like renormalization of the activation energy ${\cal E}_{0R}$, beside 
the renormalization of $V_0$. We are therefore left with the integration 
over the modes $\bar{q}_0$ and $\bar{q}_1$, with a renormalised potential 

\begin{eqnarray}
&&\int_0^{\tau} dt V_R(q_0,q_1(t)) = V_0\tau - V_{0R}\int_0^{\tau} dt 
\cos ( k(q_0+q_1(t)) ) \nonumber \\
&&{\simeq} V_0\tau - V_{0R}\tau J_0(2k|\bar{q}_1|)\cos ( kq_0 )  
\label{potentr}
\end{eqnarray}

\noindent
Here, $J_0$ is the Bessel function and the renormalised amplitude $V_{0R}$ is 

\begin{eqnarray}
V_{0R}= \left \{ \begin{array}{ll}
V_0 \left ( \frac{\Omega\tau}{2\pi} \right )^{-\eta_c/\eta}
& \mbox{if $\eta < \eta_c$} \\
V_0 \left ( \frac{\mu +1/n^{*}}{ \mu +1} \right )^{\eta_c/\eta}
& \mbox{if $\eta > \eta_c$}
\end{array} \right.
\label{amplitr}
\end{eqnarray}

\noindent
In Eq. (\ref{potentr}) the phase of the $\bar{q}_1$ mode has been integrated 
out, allowing us to integrate out the $\bar{q}_0$-mode exactly; this leads 
to the expression 

\begin{equation}
\frac{\Gamma}{2L^2}= Im \int_{\epsilon}^{\infty} d{\tau} {\cal N}(\tau) 
e^{-({\cal E}_{0R}^2+V_0)\tau} 
\int_0^{\infty} d|\bar{q}_1|^2 e^{-(\pi\eta-E^2\tau)|\bar{q}_1|^2} 
I_0 \left (
V_{0R}\tau J_0(2k|\bar{q}_1|) \right )
\label{rate1}
\end{equation}

\noindent
where $I_0$ is the modified Bessel function. It is clear that the singularity 
from the $\bar{q}_1$-integral occurs at $\tau=\pi\eta/E^2$; evaluating the 
normalization factor ${\cal N}(\tau)$, we finally arrive at

\begin{eqnarray}
&&\Gamma=\Gamma_0K(J) \\
\label{final}
&&K(J)=e(1+\mu) \left ( 1+\frac{\mu\Omega\eta}{8\pi^2 J^2} \right )
I_0 \left ( \frac{V_{0R}\eta}{4\pi J^2} J_0(2k{\ell}_N) \right ) \nonumber
\end{eqnarray}

\noindent
where $\Gamma_0$ is given by Eq. (\ref{rate0}), there is a further
renormalization ${\cal E}_{0R}^2{\rightarrow}{\cal E}_{0R}^2+V_0$ and 
we have set $E=2\pi J$. $\ell_N$ is a nucleation length, which is in first 
approximation given by 

\begin{equation}
{\ell}_N^2{\simeq}\frac{ {\cal E}_{0R}^2}{4\pi^2 J^2}
-\frac{V_{0R}}{4\pi^2 J^2} \left | J_0 \left ( k
\frac{ {\cal E}_{0R} } {\pi J} \right ) \right |
\label{nuclen}
\end{equation}
 
\noindent
and corresponds physically to the distance a vortex and antivortex 
pair must travel to acquire the nucleation energy ${\cal E}_{0R}$.
The presence of the $J_0(2k{\ell}_N)$ argument in the correction factor 
$K(J)$ due to the pinning lattice thus gives rise to oscillations in  
$\Gamma (J)$ (hence in the sample's resistance) through the parameter 
$2k{\ell}_N=4\pi{\ell}_N/d$. Vortex nucleation is therefore 
sensitive to the corrugation of the pinning substrate. However, these 
oscillations should be observable only in the confined phase, $\eta > \eta_c$,
where interrupted-renormalization prevents the prefactor in front of the
$J_0(x)$ oscillating function from becoming too small for relatively 
small current densities.

 \end{document}